\documentclass[12pt]{iopart}
\usepackage[dvips]{graphicx} 
\usepackage{amssymb}
\usepackage{epsfig}
\usepackage{color}
\usepackage{ifthen}
\usepackage{iopams}  
\usepackage{multirow} 	
\usepackage{xcolor}
\begin{document}
\def\p{\partial}
\def\half{{1\over 2}}
\def\({\left(}
\def\){\right)}
\def\[{\left[}
\def\]{\right]}
\def\be{\begin{equation}}
\def\ee{\end{equation}}
\def\beq{\begin{eqnarray}}
\def\eeq{\end{eqnarray}}
\def\nn{\nonumber}
\def\fnl{f_{\rm NL}}
\def\gnl{g_{\rm NL}}
\def\ra{\rightarrow}
\def\rr{{\bf r}}
\def\vv{{\bf v}}
\def\xx{{\bf x}}
\def\nn{{\nonumber}}
\newcommand{\dd}{\mathrm d}
\def\G{\rm G}
\title[Primordial non-Gaussian signatures in CMB polarization]{Primordial non-Gaussian signatures in CMB polarization}
\author{Vidhya Ganesan$^{1}$} 
\ead{vidhya@iiap.res.in} 
\author{Pravabati Chingangbam$^{1}$} 
\ead{prava@iiap.res.in} 
\author{K. P. Yogendran$^{2}$} 
\ead{yogendran@iisermohali.ac.in} 
\author{Changbom Park$^{3}$} 
\ead{cbp@kias.re.kr} 
\address{$^1$ Indian Institute of Astrophysics, Koramangala II Block,
  Bangalore  560 034, India\\
$^2$ Indian Institute of Science Education and Research, Sector 81, Mohali, India\\
$^3$Korea Institute for Advanced Study, 85 Hoegiro, Dongdaemun-gu,
Seoul 130-722, Korea} 

\begin{abstract}
  We study the signatures of local type primordial non-Gaussianity, parametrized by $\fnl$, of
  scalar perturbations in CMB polarization using the probability
  distribution functions, Minkowski Functionals and Betti numbers. We
  show that the lowest order non-Gaussian deviation of the PDF of the
  total polarization intensity is at order $(\fnl\sigma)^2$. We
  calculate the non-Gaussian deviations of Minkowski Functionals and
  Betti numbers from simulated polarization maps. We find that $E$
  mode polarization provides independent and equally strong constraint on $\fnl$ as
  temperature fluctuations. The non-Gaussian signal in the total
  polarization intensity, however, is much weaker and has a relatively large cosmic variance and hence may not be useful for detecting local type
  non-Gaussianity.

\end{abstract}
\maketitle

\section{Introduction} 


The Cosmic Microwave Background radiation photons are
polarized~\cite{Bond:1987ub,Crittenden:1993wm,Frewin:1994,Coulson:1994,Crittenden:1995}
at less than 10\% level due to quadrupolar anisotropies that were
present in the plasma before the epoch of matter and radiation
decoupling. The detection of this polarization is extremely valuable
and will enhance our knowledge about the universe by providing a consistency
check of the standard cosmological model that has been inferred from
temperature fluctuations and improve constraints on the cosmological
parameters. 
Inflation~\cite{Starobinsky:1979,Guth:1981,Starobinsky:1982,Linde:1982,Albrecht:1982} predicts that the amplitude and phase of the fluctuations in the  energy density during the very early stages of the Universe are random variables with a nearly Gaussian 
probability distribution function. Their precise statistical nature,
in particular the manner of deviation from Gaussianity, has been the
subject of intense study during the past decade.  All models of
inflation, in general, predict some amount of deviation of these
fluctuations from a Gaussian distribution. The details of the
deviations are model dependent making it possible to
discriminate different theoretical models using present day observations.  

The statistical nature of the primordial fluctuations are imprinted on the
fluctuations in the CMB temperature and polarization, and matter
distributions. 
Observations of these fluctuations enable us to reconstruct features of the
primordial fluctuations.  Searches for primordial non-Gaussianity
 have so far focused on using the fluctuations in the
CMB temperature. Some of these searches use geometrical and topological
observables associated with excursion sets. Of these, probably the most widely used are the Minkowski
Functionals~\cite{Tomita:1986,Gott:1990,Winitzki:1997jj,Matsubara:2003yt,Hikage:2006fe,Matsubara:2010te}.  
They have been applied to observational data of temperature fluctuations to constrain primordial non-Gaussianity ~\cite{Hikage:2008gy,Hikage:2012bs,Komatsu:2011,Planck:2013} and also to identify traces of residual foreground  contamination~\cite{Chingangbam:2012wp}. Other such observables include clustering strength of the excursion set~\cite{Rossi:2009,Rossi:2010hu,Rossi:2010},  number counts of hot and cold spots (or Betti
numbers)~\cite{Coles:1987,Weygaert:2010,Weygaert:2011a,Weygaert:2011b,Chingangbam:2012km,Park:2013dga} and extrema counts~\cite{Pogosyan:2009}. 
Studies on Betti numbers and extrema counts for the CMB have been theoretical nature and have not yet been applied to observational data. 

In this work, we focus on the so-called local type primordial
non-Gaussianity which is parametrized by the parameter
$\fnl$~\cite{salopek:1990,raghu:1991,Acquaviva:2002ud,maldacena:2002}.
We extend the study of Minkowski Functionals and Betti numbers to the
polarization field. Previous analysis of polarization along these
lines can be found in~\cite{Changbom_Changyung_1,Changbom_Changyung_2}. We calculate the
effects of primordial non-Gaussianity on the polarization signal and
compare them with what is obtained for the temperature
fluctuations focusing on scalar density fluctuations generated during
inflation. 
From temperature data, the current constraints on $\fnl$ given by PLANCK
data~\cite{Planck:2013} is $\fnl = 2.7\pm 5.8$ at 68\% CL.  Recently, Galli et. al.~\cite{Galli:2014kla} showed that
polarization data can vastly improve constraints on cosmological
parameters in comparison to using temperature data alone. Our work in
this paper is along similar lines and we investigate the shapes and
amplitudes of non-Gaussian deviations that show up in Minkowski
Functionals and Betti numbers for the polarization fields, compare
them with deviations seen for temperature fluctuations and comment on
their power to constrain $\fnl$. We do not address real observational
issues such as instrument noise, incomplete sky, beam shapes, etc. and study the non-Gaussian signal relative to the cosmic variance.

This paper is organized as follows. In section 2, we briefly describe
local type primordial non-Gaussianity and non-Gaussian simulations. In
section 3 we present analytic expressions and numerical calculations
from simulations of the Gaussian and non-Gaussian probability
distribution functions of $E$ and $I$. In section 4, we describe
Minkowski Functionals and Betti numbers and their numerical
calculation. Then we present our results for non-Gaussian deviations of
these observables. Further, we perform a simple comparison of the
statistical sensitivity of the primordial non-Gaussian information
encoded in temperature fluctuations and polarization using the
deviations of Minkowski Functionals.  We end by summarizing the results along with a discussion of their implications in section 5.
\section{Local type primordial non-Gaussianity, polarization fields and simulations}

In the local type primordial
non-Gaussianity the primordial gravitational
potential $\Phi$ takes the form
\begin{equation}
\label{eqn:phi} 
\Phi(\xx) = \Phi^G(\xx)+ \fnl  \left\{ (\Phi^G(\xx))^2 - \langle (\Phi^G)^2\rangle \right\},
\end{equation}
where $\Phi^G$ is Gaussian and $\fnl$ is a constant parameter
that quantifies the extent of non-Gaussianity.  The expression is
meaningful for values of $\fnl$ small enough so that the second term
is smaller compared to the first. $\Phi$ sets the initial  conditions in the theoretical calculation of temperature fluctuations and polarization. 

The two degrees of freedom of polarization are encoded in the Stokes
parameters $Q$ and $U$. These transform as spin 2 objects under rotations along the line of sight. They can be re-expressed in terms of the curl-free and divergence free components, the so-called $E$ and $B$ modes~\cite{Kamionkowski:1996ks,Zaldarriaga:1996xe}. $E$ is a scalar and $B$ is a pseudo-scalar. $E$ modes have been observationally detected by DASI~\cite{Kovac:2002fg} and subsequently by WMAP~\cite{Kogut:2003et}. For our analysis in this paper we use the $E-$ mode and the total polarization intensity, defined as 
$I\equiv\sqrt{Q^2+U^2}$. In general, $Q$ and $U$ contain independent information. Here we consider the case where $B$ mode is absent (the tensor-to-scalar ratio, $r$ being zero). Under this  condition $Q$ and $U$ are correlated and $E$ mode contains the full information in the two modes. 

We use simulations of temperature fluctuations and $E$ mode
polarization with input primordial fluctuations of the form given by
Eq.~(\ref{eqn:phi}) that have been made publicly available by Elsner
and Wandelt~\cite{Elsner:2009md}.  The simulations follow the
algorithm given in~\cite{Liguori:2003} and involve calculating
$a^i_{\ell m}$ where $i$ can be either $\Delta T/T_{\rm CMB}$ or $E$.
$a^i_{\ell m}$ is
obtained from
\begin{eqnarray}
a^i_{\ell m} &=& \int \dd r \,r^2\, \Phi_{\ell m}(r)\,
\Delta^i_{\ell}(r) \nn \\ &=& \int \dd r\, r^2\,\Phi^{\rm G}_{\ell
  m}(r)\, \Delta^i_{\ell}(r) + \int \dd r\, r^2\,\Phi^{\rm NG}_{\ell
  m}(r)\, \Delta^i_{\ell}(r),
\end{eqnarray}
where $\Delta^i_{\ell}(r)$ is the transfer function for the respective
$i$, and $\Phi_{\ell m}(r)$ is the harmonic transform of $\Phi({\bf
  x})$ with Gaussian and non-Gaussian parts denoted by $\Phi^{\rm G}$
and $\Phi^{\rm NG}$.  For expressions relating $\Phi({\bf x})$ and
$\Phi_{\ell m}(r)$ we refer to~\cite{Liguori:2003}. The fields $\Delta T/T_{\rm CMB}$ or $E$ are obtained by performing 
harmonic transform of the respective $a^i_{\ell m}$'s. 
A similar 
algorithm was applied to generate maps of temperature fluctuations for cubic order
perturbations of $\Phi$~\cite{Chingangbam:2009vi}. The differences in the maps of $\Delta T/T_{\rm CMB}$ and $E$ arise from different physical effects encoded in their respective transfer functions.  

The simulations have resolution set by the maximum multipole
$\ell_{max}=1024$ and the HEALPIX variable $n_{side}=512$.  The input
cosmological parameters are those obtained from WMAP5+BAO+SN data
given in~\cite{wmap5}. 
We consider Gaussian smoothing of the input
maps to study the non-Gaussian effects at different resolutions
parametrized by smoothing angles $\theta_s$. Note that $\theta_s$ is
related to the FWHM as $\theta_s= {\rm FWHM}/\sqrt{8\ln 2}$. 

For the calculations of Minkowski Functionals and Betti numbers we
use the mean shifted total polarization intensity denoted by
\begin{equation}
{\tilde I} \equiv I-\langle I\rangle.
\end{equation}

\section{Probability distribution functions and non-Gaussian
  deviations}

The statistical properties of the primordial gravitational potential
are directly reflected in $\Delta T/T_{\rm CMB}$ and $E$ provided the
perturbations evolve linearly during subsequent epochs. This implies 
that if the probability distribution function (PDF) of $\Phi$ is
Gaussian then the PDF of $\Delta T/T_{\rm CMB}$ and $E$ will also be
Gaussian~\cite{Bond:1987ub}. On the other hand if the PDF has a 
non-Gaussian form then $\Delta T/T_{\rm CMB}$ and $E$ will trace the same
distribution. Similarly, $Q$ and $U$ will trace the PDF of $\Phi$, which in turn will lead to the PDF for $I$. 

Let P denote the PDF for a generic field. 
We first derive P, for a field which
has the form of Eq.~(\ref{eqn:phi}) and whose Gaussian part is denoted by $X$. 
The expectation value of a function
$f\left(X+\fnl \left( X^2-\sigma^2  \right)\right)$, where $\sigma^2=\langle X^2\rangle$, may be computed using the Fourier transform as
$$
\langle f\rangle = \int \dd X \,{\rm P}(X) f(X+\fnl (X^2-\sigma^2)) 
= \int \dd X\, {\rm P}(X) \int \frac{\dd k}{\sqrt{2\pi}}
e^{i k(X+\fnl (X^2-\sigma^2))} \tilde f(k).\nn
$$
Expanding the right hand side of the above in a series in $\fnl$, and performing a double Fourier transform, we can write
$$ 
\langle f\rangle =\int \dd X f(X) \left( {\rm P}^{(0)}(X)+{\rm P}^{(1)}(X)+ \ldots\right)
$$ 
The first term is the expectation value in the absence of $\fnl$ given by
$P^{(0)}(X)=\frac{1}{\sqrt{2\pi\sigma^2}}\exp(-\frac{X^2}{2\sigma^2})$, while the
higher order terms may be interpreted as ``corrections'' to the
Gaussian distribution. The first higher order term is obtained as
\begin{equation}
{\rm P}^{(1)}(X)= -\frac{\fnl\sigma}{\sqrt{2\pi\sigma^2}} \,  e^{-X^2/2\sigma^2} \,\frac{X(X^2-3\sigma^2)}{\sigma^3}
\label{eqn:dp1}
\end{equation}
and the next order is 
\begin{equation}
{\rm P}^{(2)}=\frac{(\fnl\sigma)^2}{\sqrt{2\pi\sigma^2}}
 \,e^{-X^2/2\sigma^2}\,
\frac{(X^6-11X^4\sigma^2+23 X^2\sigma^4 -5\sigma^6)}{2\sigma^6}
\label{eqn:dp2}
\end{equation}

\begin{figure}[h]
\begin{center}
\resizebox{3in}{3.in}{\includegraphics{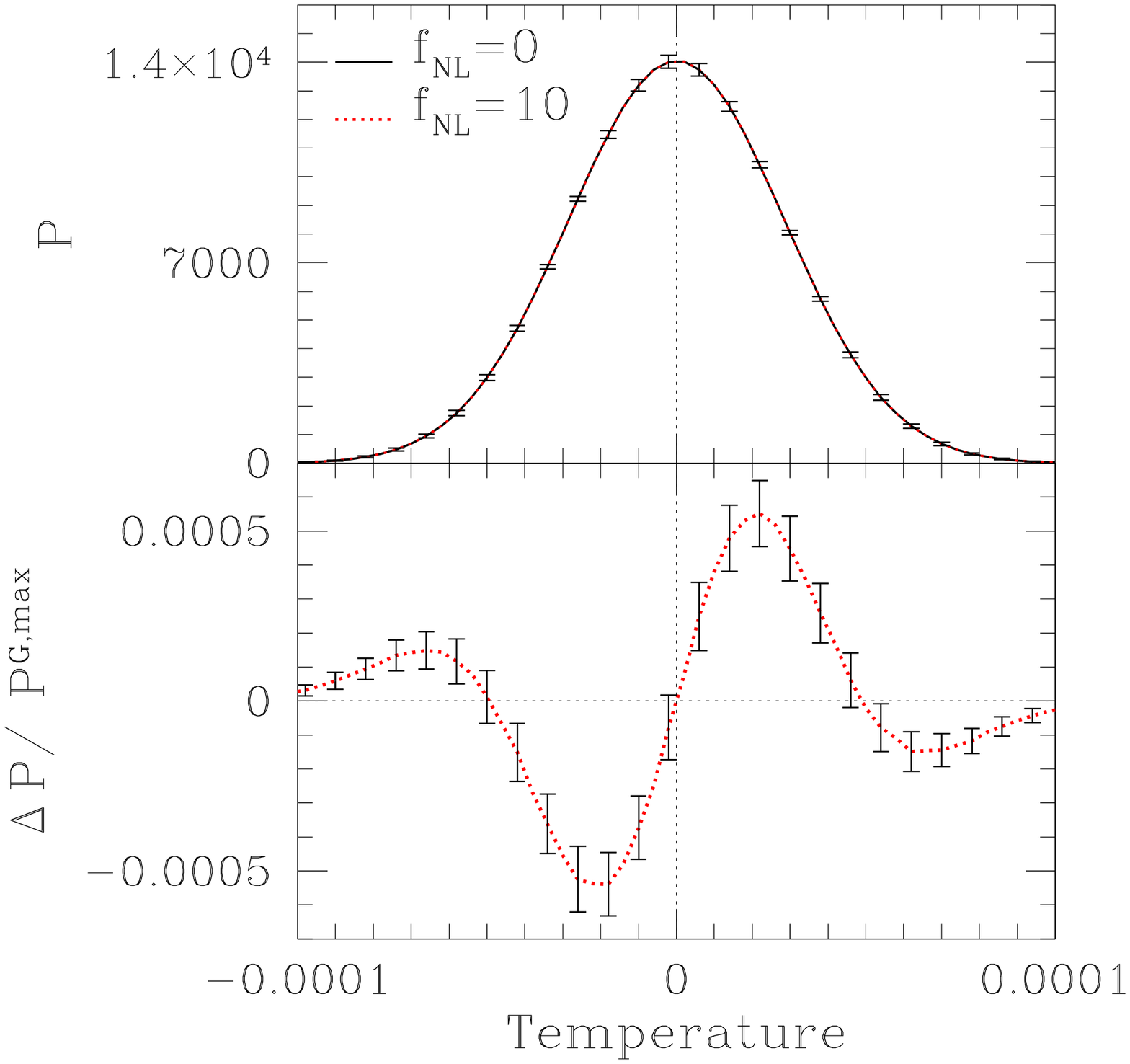}}
\resizebox{3in}{3.in}{\includegraphics{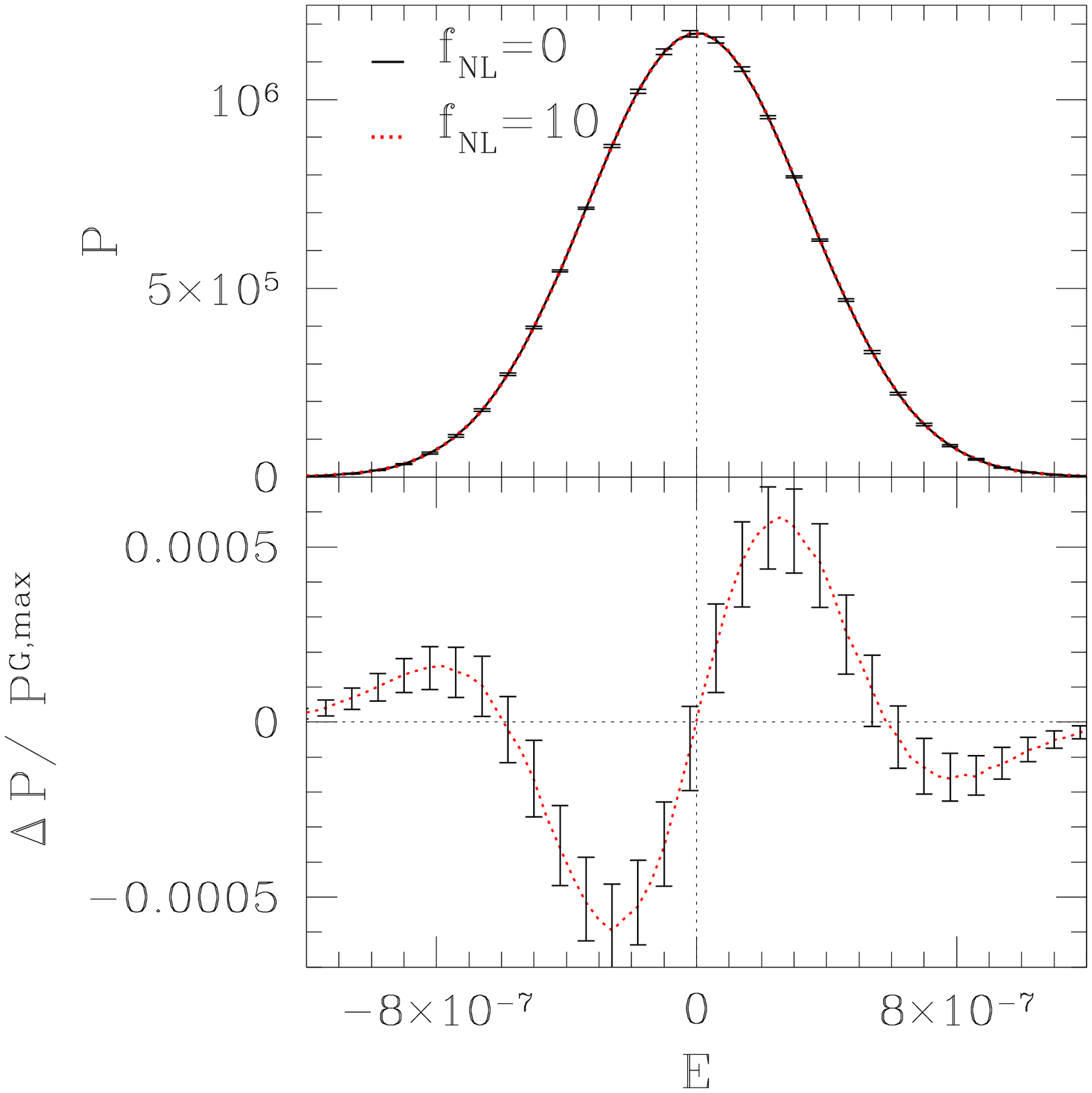}}
\end{center}
\caption{ PDF and its non-Gaussian deviation for temperature
  fluctuations ({\em left panel}) and $E$ mode ({\em right panel}) for
  smoothing angle $\theta_s=20'$. Plots are average over 1000
  simulations. Error bars are the sample variance from the 1000
  simulations.}
\label{fig:pdf_TE}
\end{figure}

We expect the PDF of $\Delta T$ and $E$ to have non-Gaussian deviation
of the form given by Eq.~(\ref{eqn:dp2}).  The amplitude and the rms
however will be modified by the physical events during recombination
and subsequent epochs.  In the left and middle panels of
Fig.~(\ref{fig:pdf_TE}) we have plotted the normalized PDF for
temperature fluctuations and $E$ mode, averaged over 1000 simulations. Upper panels show the PDF for both Gaussian and
non-Gaussian cases, while lower panels show $\Delta \,{\rm P}$ scaled
by the amplitude of ${\rm P}^{(0)}$, for $\fnl =10$. We have used
smoothing angle $\theta_s=20'$. We find that the differences between the Gaussian and non-Gaussian maps in shapes and amplitudes
for the two fields are similar, as seen from the lower panels.  The shapes are well approximated by Eq.~(\ref{eqn:dp1}).  The second order correction
Eq.~(\ref{eqn:dp2}) is negligible for such small value of $\fnl$.

In Fig.~(\ref{fig:pdf_TE}) the non-Gaussian deviation has been calculated between each pair of Gaussian and non-Gaussian maps having the same set of Gaussian $a_{\ell m}$'s. Therefore, a large part of the cosmic variance associated with the fluctuations in the amplitude of harmonic modes are cancelled out and only the part associated with the higher order term contributes to the sample variances shown in the bottom panel of  Fig.~(\ref{fig:pdf_TE}). This treatment might look unrealistic for a map obtained from real observation because the corresponding Gaussian map is not known. However, in such a situation the Gaussian map can be created by randomizing the phase of harmonic modes of the observed map while keeping the amplitude of the $a_{\ell m}$'s the same.  This recipe can reduce the cosmic variance in the non-Gaussian deviation associated with random fluctuation of the shape of the power spectrum and make the deviation much more sensitive to true non-Gaussianity. In the next section the non-Gaussian deviations of Minkowski Functionals and Betti numbers are calculated in the same way. Hence
the comments made above will be relevant there too. 

To calculate the PDF of the total polarization intensity we consider a field of the form $R\equiv \sqrt{X^2+Y^2}$,
where $X$ and $Y$ are independent random variables. We start
with the definition
\begin{equation}
{\rm P}(R,\theta)\,\dd R\,\dd\theta\equiv
{\rm P}(X)\,{\rm P}(Y)\,\dd X\,\, \dd Y
\end{equation}
where $X=R\cos\theta,\ Y=R\sin\theta$, and integrate over
$\theta$. For Gaussian $X$ and $Y$ this gives,
\begin{equation} 
{\rm P^{(0)}}(R)\ =\ \frac{R}{\sigma^2}  \, e^{-\frac{R^2}{2\sigma^2}},   
\label{eqn:pI}
\end{equation}
where the range for $R$ is $[0,\infty)$. For non-Gaussian $X$ and $Y$
of the form given by Eq.~(\ref{eqn:phi}) we can use ${\rm P}(X)={\rm
  P^{(0)}}+{\rm P}^{(1)}+{\rm P}^{(2)}+\ldots$, where ${\rm P}^{(1)}$ and
${\rm P}^{(2)}$ are given by Eqs.~(\ref{eqn:dp1}) and (\ref{eqn:dp2})
respectively, and substitute in Eq.~(6) to get the non-Gaussian
correction to the PDF for $R$ order by order in $\fnl\sigma$.
Because ${\rm P}^{(1)}(X)$ is odd in $X$, 
the PDF for $R$ receives no correction at the first
order in $\fnl\sigma$. Then to order $(\fnl\sigma)^2$ the PDF becomes 
\begin{equation}
{\rm P}(R) = \frac{R}{\sigma^2} e^{-\frac{R^2}{2\sigma^2}}
\left(1+\frac{\fnl^2\sigma^2}{16\sigma^{6}}(5R^6\,-66R^4\,\sigma^2+184\, R^2\,\sigma^4-80\,\sigma^6) +\ldots\right)
\label{eqn:dpI}
\end{equation}
Thus, we expect smaller non-Gaussian deviations to show up in $I=R$ in
comparison to $E$ (since $f_{NL}\sigma <<1$).  

\begin{figure}[h]
\begin{center}
\resizebox{2.5in}{2.5in}{\includegraphics{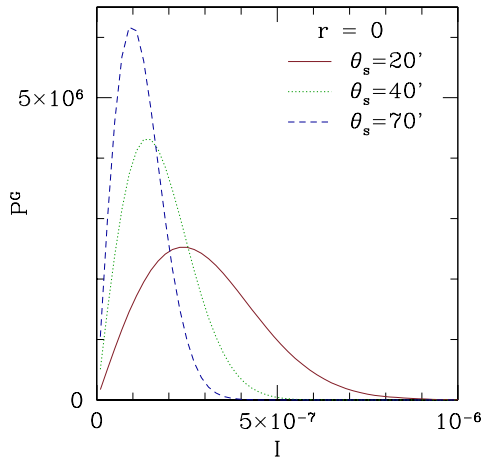}}
\end{center}
\caption{ PDF of Gaussian $I$ where the input $E$ is Gaussian and
  smoothed with different smoothing angles. }
\label{fig:pdf_I}
\end{figure}

$Q$ and $U$ are linearly independent fields and hence the PDF of $I$, when they
are Gaussian, must be of the form Eq.~(\ref{eqn:pI}). In  Fig.~(\ref{fig:pdf_I}) we have plotted the normalized PDF for
$I$ for input Gaussian $E$ for different $\theta_s$. 
The smoothing was carried out on the input $E$ and then $I$ was
constructed from the smoothed fields. The shapes are in agreement with
Eq.~(\ref{eqn:pI}). For a comparison we show the case when $B$ mode is included in the Appendix A.

\begin{figure}[h]
\begin{center}
\resizebox{3.in}{3.in}{\includegraphics{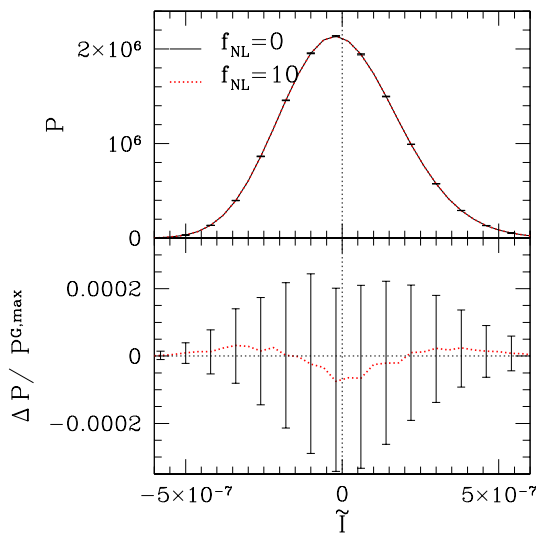}}
\end{center}
\caption{ Same as Fig.~(\ref{fig:pdf_TE}) for ${\tilde I}$ for input
  Gaussian and non-Gaussian $E$, with smoothing done on $I$. }
\label{fig:pdf_Itilde}
\end{figure}

It is important to note that if Gaussian smoothing is carried out on
$I$ the PDF shape gets modified by the Gaussian kernel of the
smoothing and becomes closer to Gaussian shape in accordance with central limit theorem.
But this does not change the fact that the deviation is at
$(\fnl\sigma)^2$ order since smoothing is a linear process.  
In Fig.~(\ref{fig:pdf_Itilde}) we show the PDF and its non-Gaussian deviation
for the case where we have taken the simulated $E$ map and performed
the smoothing on $I$ and then subtracted the mean.  The top panel
shows the PDFs for input Gaussian and non-Gaussian $E$ which are
difficult to distinguish by eye, while the lower panel shows $\Delta
\,{\rm P}$ scaled by the amplitude of ${\rm P}^{(0)}$, for $\fnl
=10$. The amplitude of non-Gaussian deviation is about an order of
magnitude lower than that of temperature fluctuations and $E$, and
the error bars are about twice larger.
For the calculations in the remainder of this paper we will use the
variable $\tilde I$ with smoothing carried out on $I$.

\section{Geometrical and topological observables} 

\begin{figure}[h]
\begin{center}
\resizebox{2.in}{2.1in}{\includegraphics{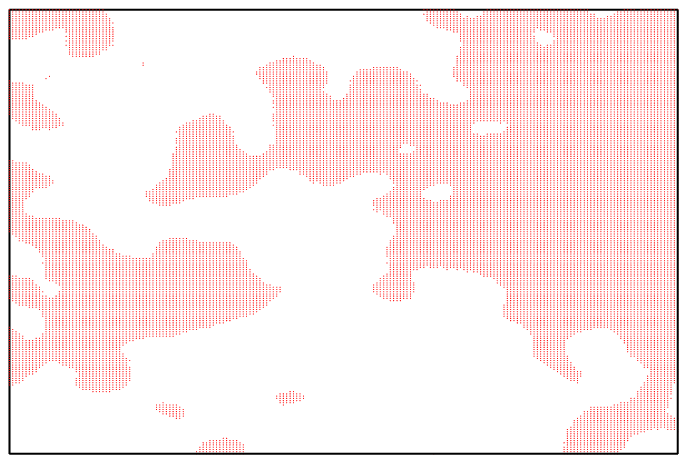}}
\resizebox{2.in}{2.1in}{\includegraphics{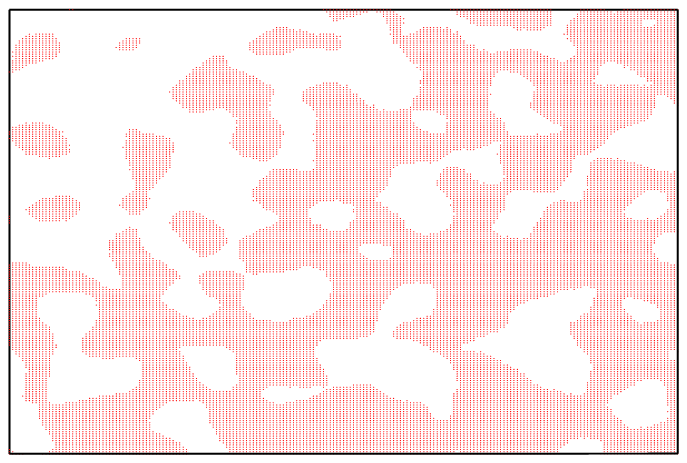}}
\resizebox{2.in}{2.1in}{\includegraphics{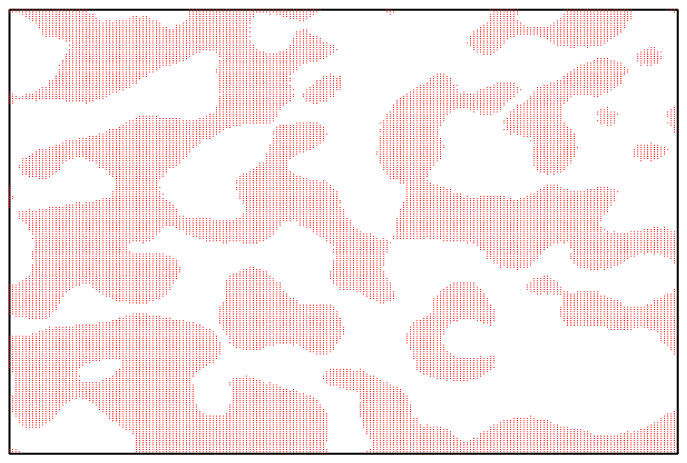}}
\end{center}
\caption{{\em Left panel}: A $7^{\circ}\times 7^{\circ}$ patch of the excursion set for a Gaussian 
  CMB temperature fluctuation field for threshold level $\nu=0$ and smoothing angle $\theta_s=90'$. {\em Middle panel}:
  The excursion set for $E$ mode for the same sky patch, Gaussian
  realization, $\nu$ and $\theta_s$.  {\em Right panel}: Same for $\tilde I$. }
\label{fig:esets}
\end{figure}

The morphological properties of excursion sets of random fields, which
is the set of all points or pixels that have values greater than or
equal to a chosen threshold value of the field, can reveal the
Gaussian or non-Gaussian nature of the fields. These properties have
systematic behavior as functions of the threshold values that
parametrize the excursion sets. We focus here on excursion sets of $E$
and $\tilde I$.  For a visual comparison, in Fig.~(\ref{fig:esets})
we show excursion sets (red contiguous regions) for a small patch of
the sky for temperature fluctuation (left panel), $E$ mode (middle
panel) and $\tilde I$ (right panel), for threshold value $\nu=0$. All
three have been obtained from the same input primordial Gaussian
fluctuation.

The morphological properties of the excursions sets can be quantified
in terms of geometrical and topological quantities namely, the
Minkowski Functionals (MFs) and Betti numbers.  There are three MFs
for two-dimensional manifolds such as the excursion sets of the
CMB. The first, denoted by $V_0$, is the area fraction of the
excursion set. The second, denoted by $V_1$, is the total length of
iso-temperature contours or boundaries of the excursion set. The
third, denoted by $V_2$, is the genus which is the difference between
the numbers of hot spots and cold spots. For two-dimensional manifolds
there are two non-zero Betti numbers, the first, denoted by
$\beta_0$, is the number of connected components, while the second
denoted by $\beta_1$ is the number of holes. (Strictly speaking, for fields on the surface of a sphere $\beta_1$ is the number of holes minus one). For the CMB excursion
sets $\beta_0$ is the number of hot spots and $\beta_1$ is the number
of cold spots.

For a Gaussian random field the MFs are given by, 
\begin{equation}
\label{eq:mf}
V_k(\nu)=A_k \, H_{k-1}(\nu)\,e^{-\nu^2/2}, \quad k=0,1,2,
\end{equation}
where $H_k(\nu)$ is the $k$-th Hermite polynomial and the amplitude $A_k$
depends only on the angular power spectrum 
$C_{\ell}$. It is given by 
\begin{equation}
\label{eq:ak}
A_k=\frac1{(2\pi)^{(k+1)/2}}\frac{\omega_2}{\omega_{2-k}\omega_k}
\left(\frac{\sigma_1}{\sqrt{2}\sigma_0}\right)^k,
\end{equation}
\begin{equation}
\label{eq:vk}
\sigma_j^2\equiv \frac1{4\pi}\sum_{\ell}(2\ell+1)\left[\ell(\ell+1)\right]^j C_{\ell}
W^2_{\ell}, 
\end{equation}
with $\omega_k\equiv \pi^{k/2}/{\Gamma(k/2+1)}$.  $\sigma_1$ is the
rms of the gradient of the field and $W_{\ell}$ represents the
smoothing kernel determined by the pixel and beam window functions and
any additional smoothing. For Gaussian smoothing $W_{\ell}$ is given
by $W_{\ell} = e^{-\ell(\ell-1)\theta_s^2/2}$.

The presence of any small deviation from Gaussianity will
appear as deviations from these Gaussian
formulae~\cite{Matsubara:2003yt, Hikage:2006fe,Matsubara:2010te}.
We denote the non-Gaussian deviation of MFs by 
\begin{equation}
\Delta V_i \equiv V_i^{NG} - V_i^{G}, 
\label{eqn:dmf}
\end{equation}
where $i=0,1,2$.

Analytic expressions for Betti numbers are not known even for Gaussian
fields. They can be formally expressed as
\begin{equation}
 \beta_0 = \frac{1}{2\pi}\int_{C_+} K \,\dd s, \quad \beta_1 = \frac{1}{2\pi}\int_{C_-} K \,\dd s,
\label{eqn:bettiformula}
\end{equation}
where $K$ is the total curvature of iso-temperature contours for each $\nu$. $C_+$ denotes contours that enclose hot spots while $C_-$ denotes contours that enclose cold spots. Their non-Gaussian deviations are denoted as 
\begin{equation}
\Delta \beta_i \equiv \beta_i^{NG} - \beta_i^{G},
\label{eqn:dbeta}
\end{equation}
where $i=0,1$ for Betti numbers.

\subsection{Non-Gaussian deviations of Minkowski Functionals}
\begin{figure}[h]
\begin{center}
\resizebox{5.in}{3.6in}{\includegraphics{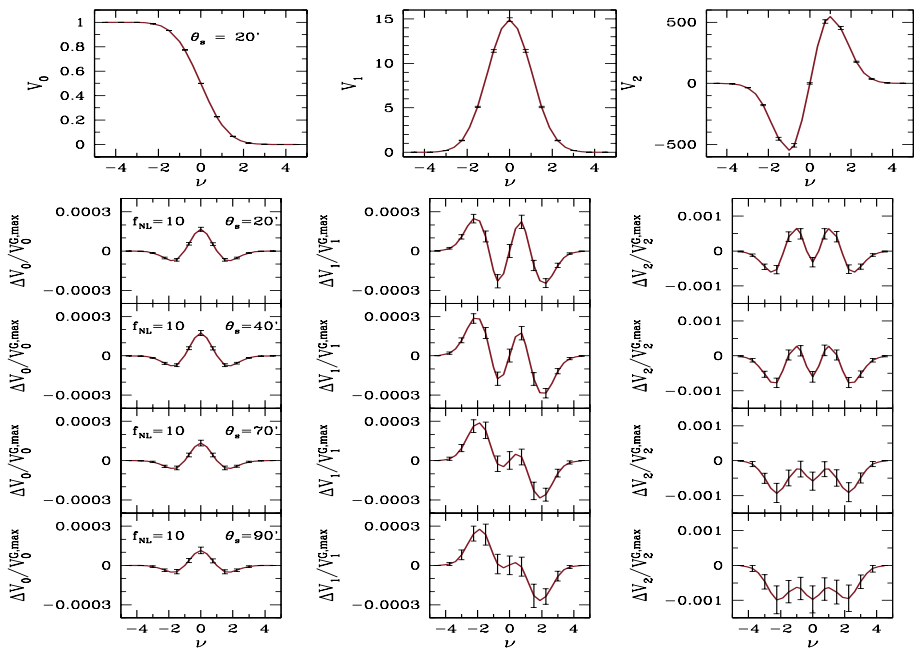}}
\end{center}
\caption{{\em Upper panels}: MFs for $E$ mode. {\em Lower panels}:
  Non-Gaussian deviations of MFs for $\fnl=10$ for different smoothing
  scales. The error bars are the sample variance from 1000
  simulations.}
\label{fig:dmfemode}
\end{figure}
We have calculated MFs numerically following the method given
in~\cite{Schmalzing:1997uc}. It was shown in~\cite{Lim:2012} that this
method has an inherent numerical inaccuracy which scales as the square
of the finite binning of the threshold values at leading order,
arising from the finite approximation of the delta function. This
issue is not of concern here for the following reason. For very weakly
non-Gaussian case, which is relevant here, the numerical errors of
Gaussian and non-Gaussian MFs for the same Gaussian realization are
similar and are subtracted off when we calculate the non-Gaussian
deviations. We have chosen spacing of $\nu$ given by $\Delta\nu=0.75$
for the range $-4.5 \le \nu \le 4.5$.

In the three upper panels of Fig.~(\ref{fig:dmfemode}) we have plotted
the MFs and their non-Gaussian deviations for $E$ mode for $\fnl = 10$
for different smoothing angles. Note that the amplitudes scale
linearly with $\fnl$. The plots are the average over calculations from
1000 simulations and the error bars are the sample variance over those
1000 simulations. As seen in the panels showing $V_1$ and $V_2$, the
contour length and the genus for $E$ have much larger amplitudes in
comparison to that of temperature fluctuations (see Fig.~(2) of
\cite{Hikage:2006fe}). This indicates more structure, which may be
guessed by visual inspection of Fig.~(\ref{fig:esets}). The
non-Gaussian deviations, shown in the lower panels, have
characteristic shapes and vary slightly with the smoothing angle.  The
error bars increase in size as $\theta_s$ increases due to loss of
statistical significance arising from fewer number of structures.
Further, we find that deviations for all three MFs are similar in
shape to that of temperature fluctuations and of comparable
amplitude. The sizes of error bars are also similar.  This implies
that $E$ modes carry as much information about non-Gaussian deviations
as the temperature fluctuations and hence can be very useful for
constraining primordial non-Gaussianity.

\begin{figure}[h]
\begin{center}
\resizebox{5.in}{3.6in}{\includegraphics{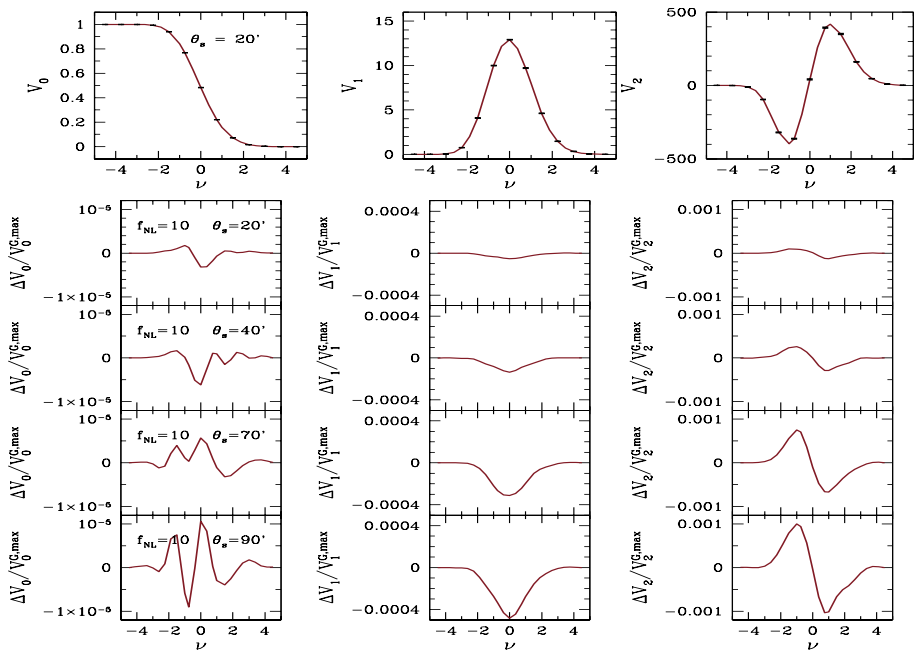}}
\end{center}
\caption{{\em Upper panels}: MFs for $\tilde I$. {\em Lower panels}:
  Non-Gaussian deviations of MFs for $\fnl=10$ for different smoothing
  scales. }
\label{fig:dmfintensity}
\end{figure}
\begin{figure}[h]
\begin{center}
\resizebox{5.in}{3.6in}{\includegraphics{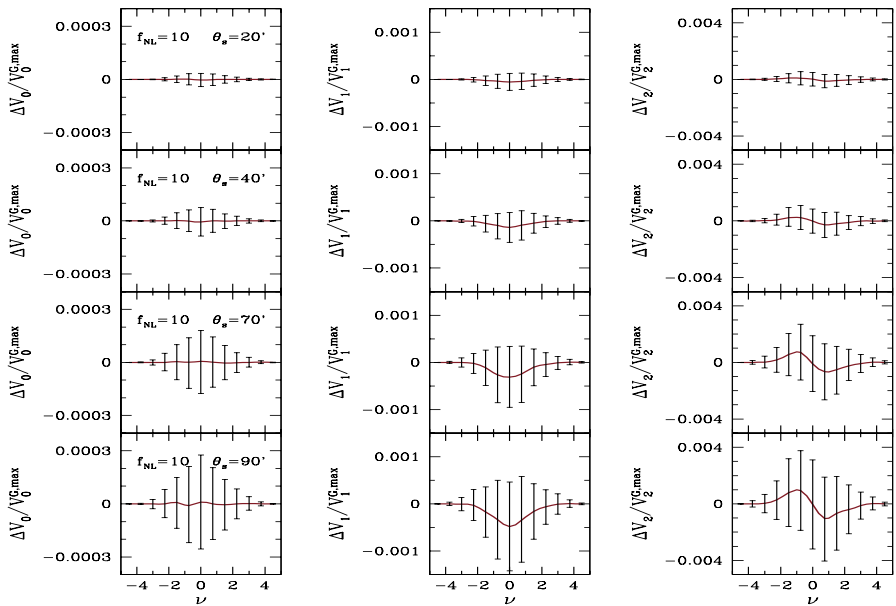}}
\end{center}
\caption{{\em Upper panels}: Non-Gaussian deviations of MFs for
  $\tilde I$ shown with error bars for the same $\fnl$ value and
  smoothing angles as lower panels of Fig.~(\ref{fig:dmfintensity}),
  but with larger scale to highlight the large size of the error
  bars. The error bars are the sample variance from 1000 simulations.}
\label{fig:dmfintensity_errorbar}
\end{figure}

Next we look at the MFs for $\tilde I$. In
Fig.~(\ref{fig:dmfintensity}) we have plotted the MFs (upper panels)
and their non-Gaussian deviations for $\fnl = 10$ for different
smoothing angles without error bars (middle panels), and the same with
error bars (lower panels). The plots are again average over
calculations from 1000 simulations and the error bars have been
obtained from the same 1000 simulations. The top panels show that the
shapes are close to that of Gaussian shapes even though $\tilde I$ is
not a Gaussian field. The reason is the additional Gaussian smoothing,
as discussed in section (3). The amplitude is close to that of $E$
modes at the same smoothing angle. From the middle panels, we see that
the shapes of deviations are quite different from that of $E$. At
lower smoothing angles, the deviations amplitudes are much smaller
than that of $E$, and become comparable for the larger ones.
Fig.~(\ref{fig:dmfintensity_errorbar}) shows the same deviations as
Fig.~(\ref{fig:dmfintensity}) but on a larger scale in order to
highlight the large error bars. This implies that statistical
fluctuations for non-Gaussian deviations of $\tilde I$ are larger
than that of $E$ or temperature fluctuations and consequently it has
considerably less power to detect local type non-Gaussian
model. 

\subsection{Non-Gaussian deviations of Betti numbers}

For the calculation of Betti numbers we follow the method given
in~\cite{Gott:1990,Chingangbam:2012km}. This method uses a numerical implementation of Eq.~(\ref{eqn:bettiformula}). It is based on connecting iso-temperature  pixels with the information of 
the orientation of the contour retained.  Contours with
the same orientation are then counted to get $\beta_0$ and  $\beta_1$.  
We would like to mention that we have not analyzed the accurarcy of this technique to the level required by present day resolutions. Hence, though our results are correct qualitatively, we do not use them for accurate statistical analysis or apply it to observational data as yet.  

\begin{figure}[h]
\begin{center}
\resizebox{1.8in}{1.8in}{\includegraphics{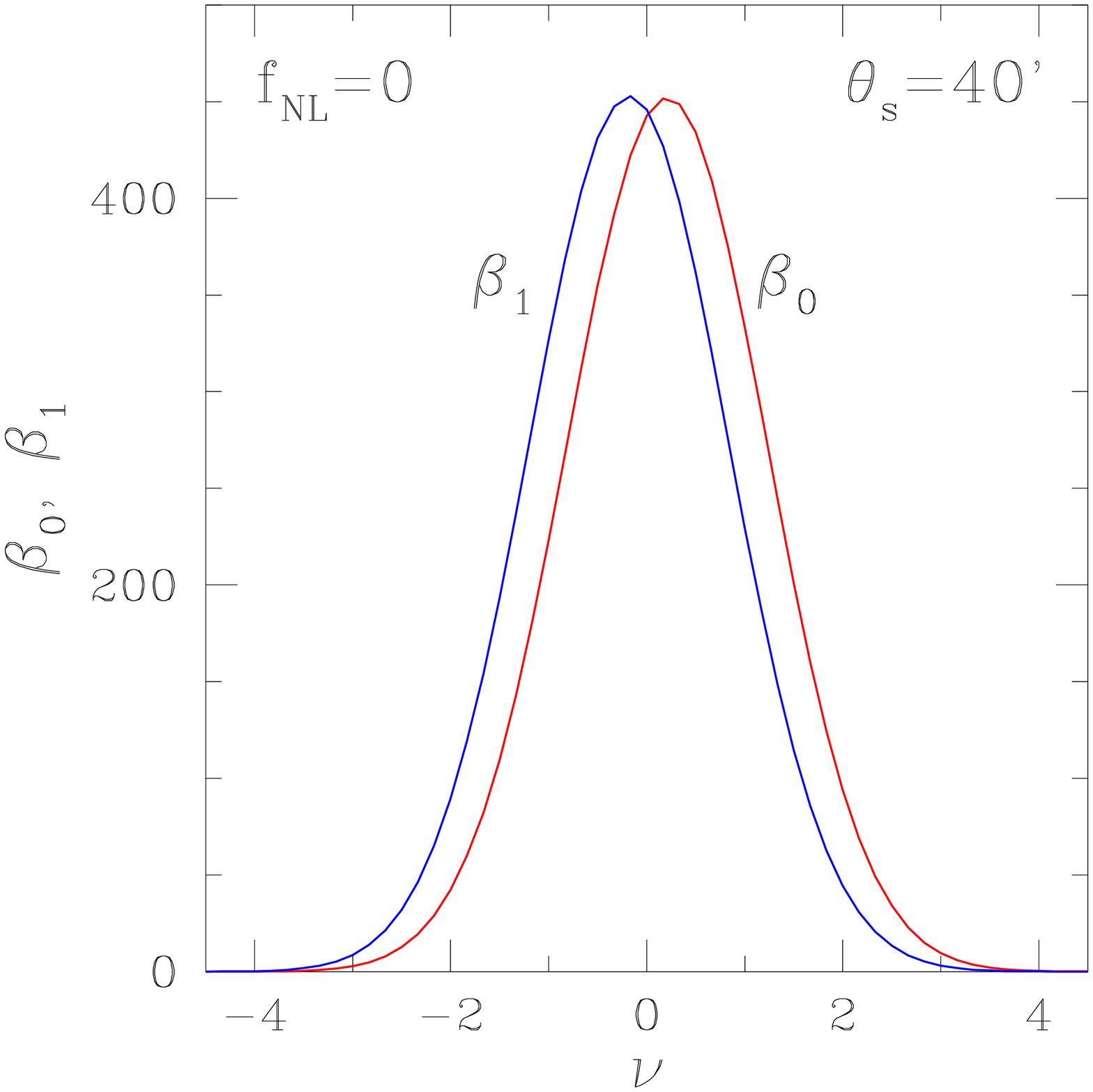}}
\resizebox{1.8in}{1.8in}{\includegraphics{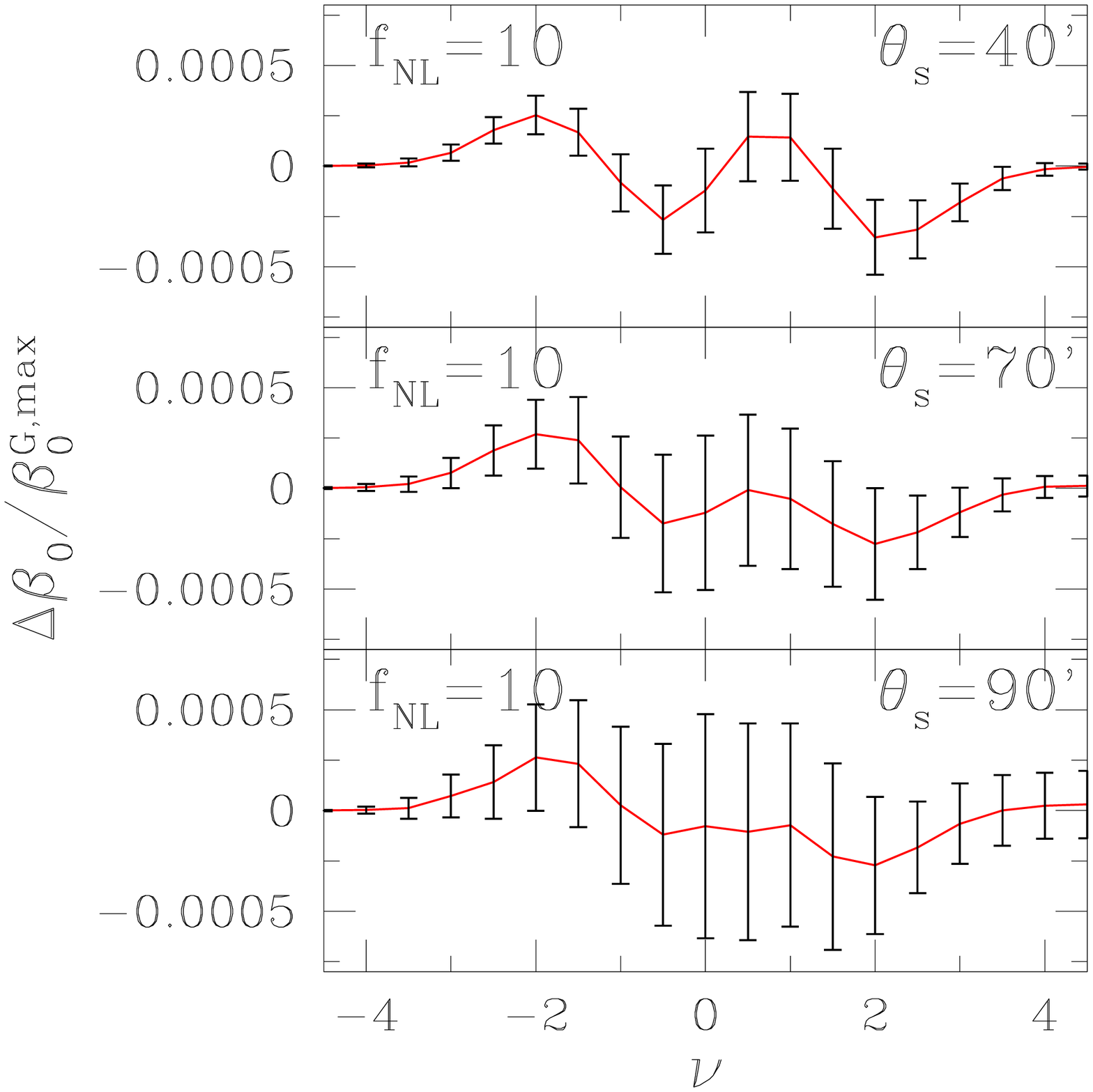}}
\resizebox{1.8in}{1.8in}{\includegraphics{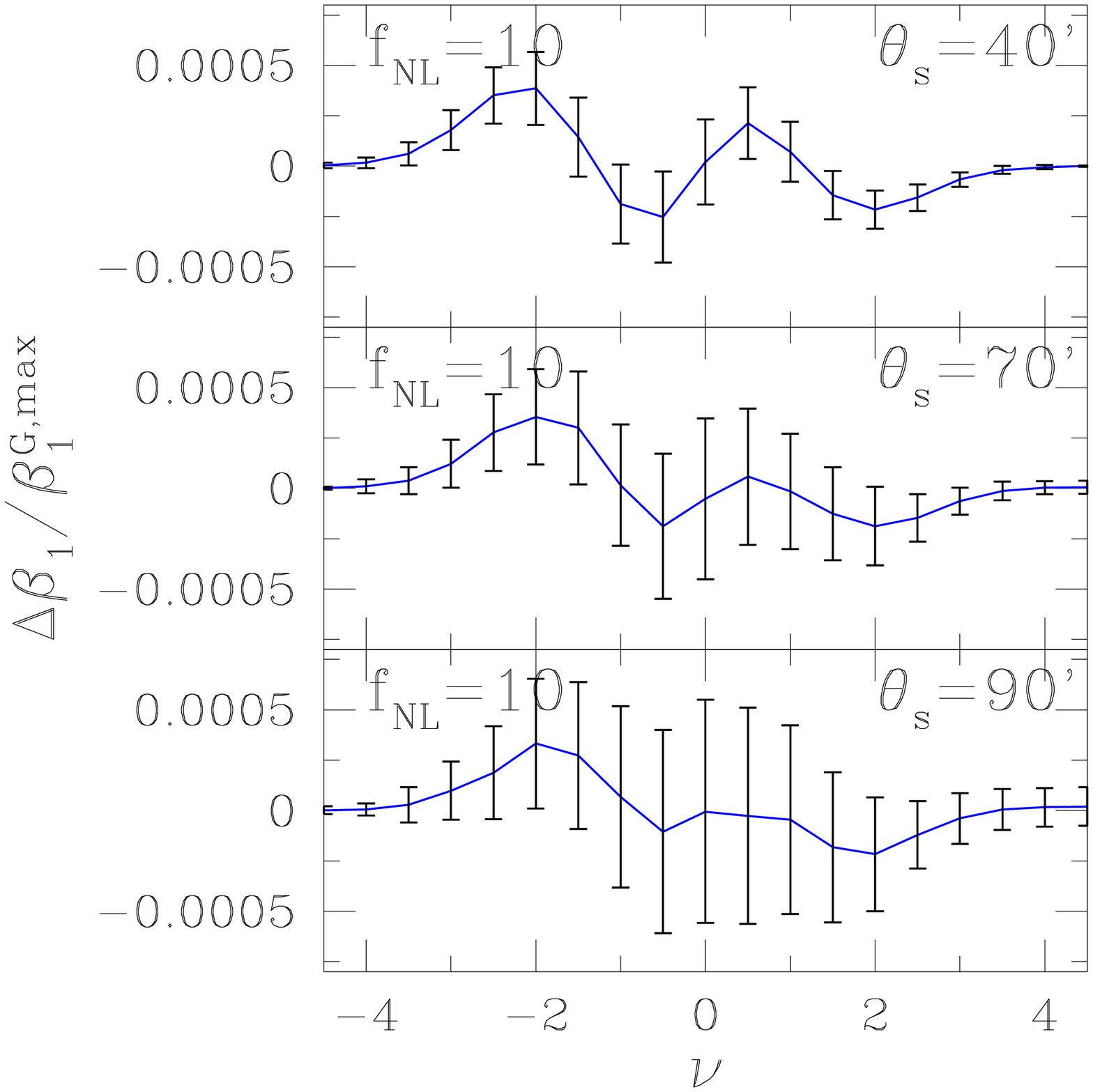}}
\end{center}
\caption{Betti numbers and their non-Gaussian deviations for $E$, for
    $\fnl=10$ and different smoothing angles.}
\label{fig:bettiemode}
\end{figure}

We have used spacing of $\nu$
given by $\Delta\nu=0.5$ for the range $-4.5 \le \nu \le
4.5$. Fig.~(\ref{fig:bettiemode}) shows $\beta_0$ and $\beta_1$ for
Gaussian case (left panel) and their respective non-Gaussian
deviations (middle and right panels) for $E$, averaged over 1000
simulations. The deviation shapes, amplitudes and sizes of error bars
are again comparable to that of temperature fluctuations (see
Figs.~(3) and (8) of~\cite{Chingangbam:2012km}).

\begin{figure}[h]
\begin{center}
\resizebox{1.8in}{1.8in}{\includegraphics{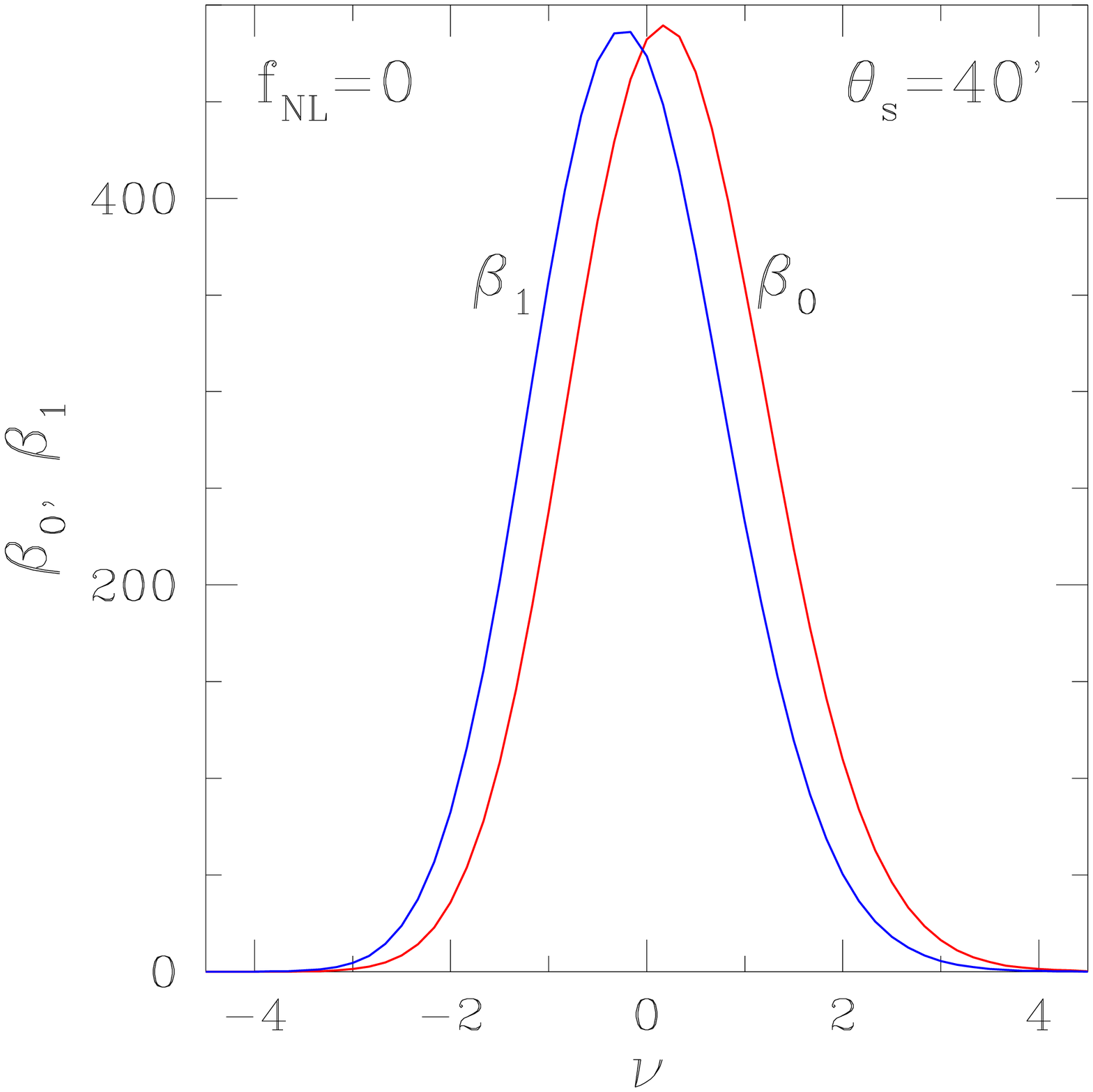}}
\resizebox{1.8in}{1.8in}{\includegraphics{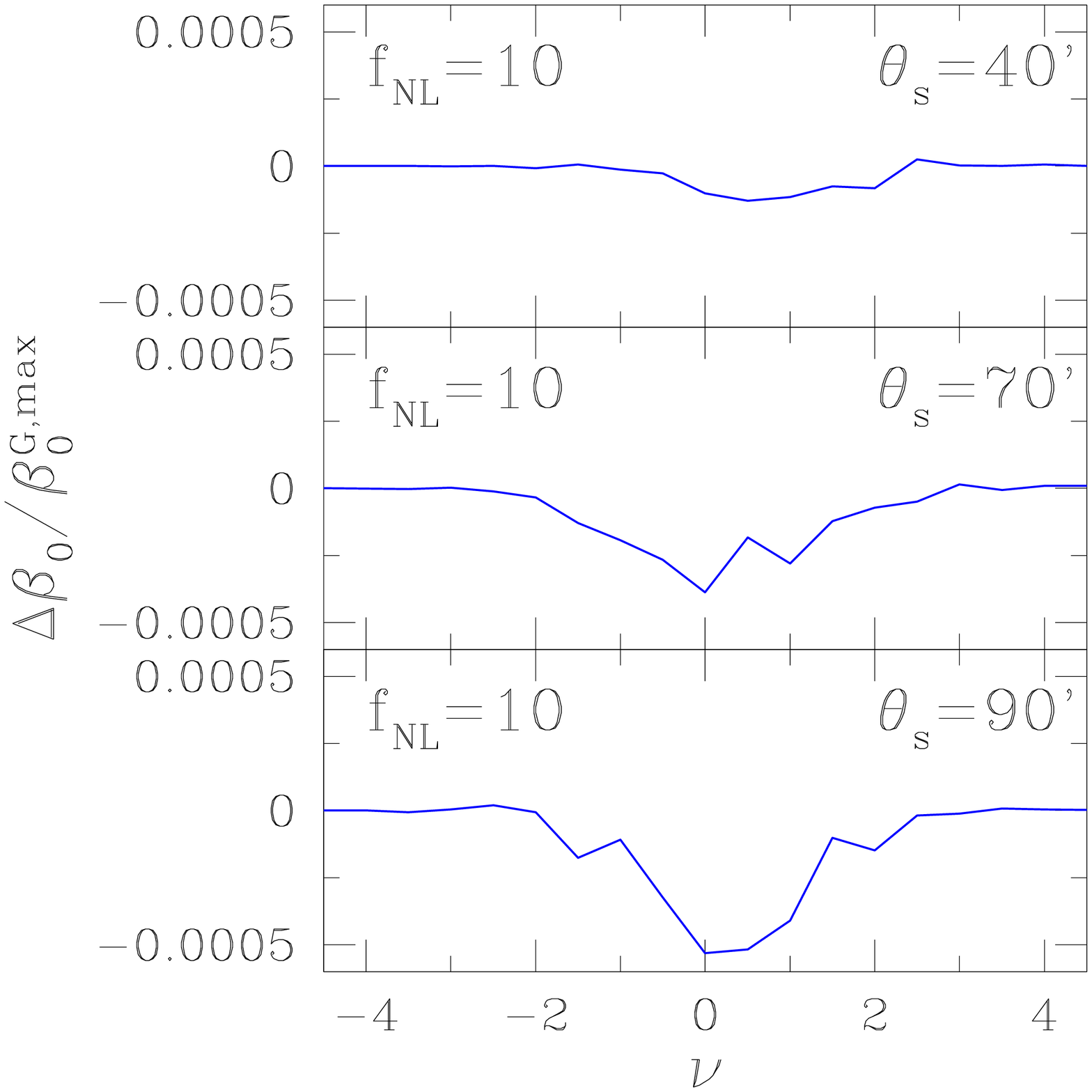}}
\resizebox{1.8in}{1.8in}{\includegraphics{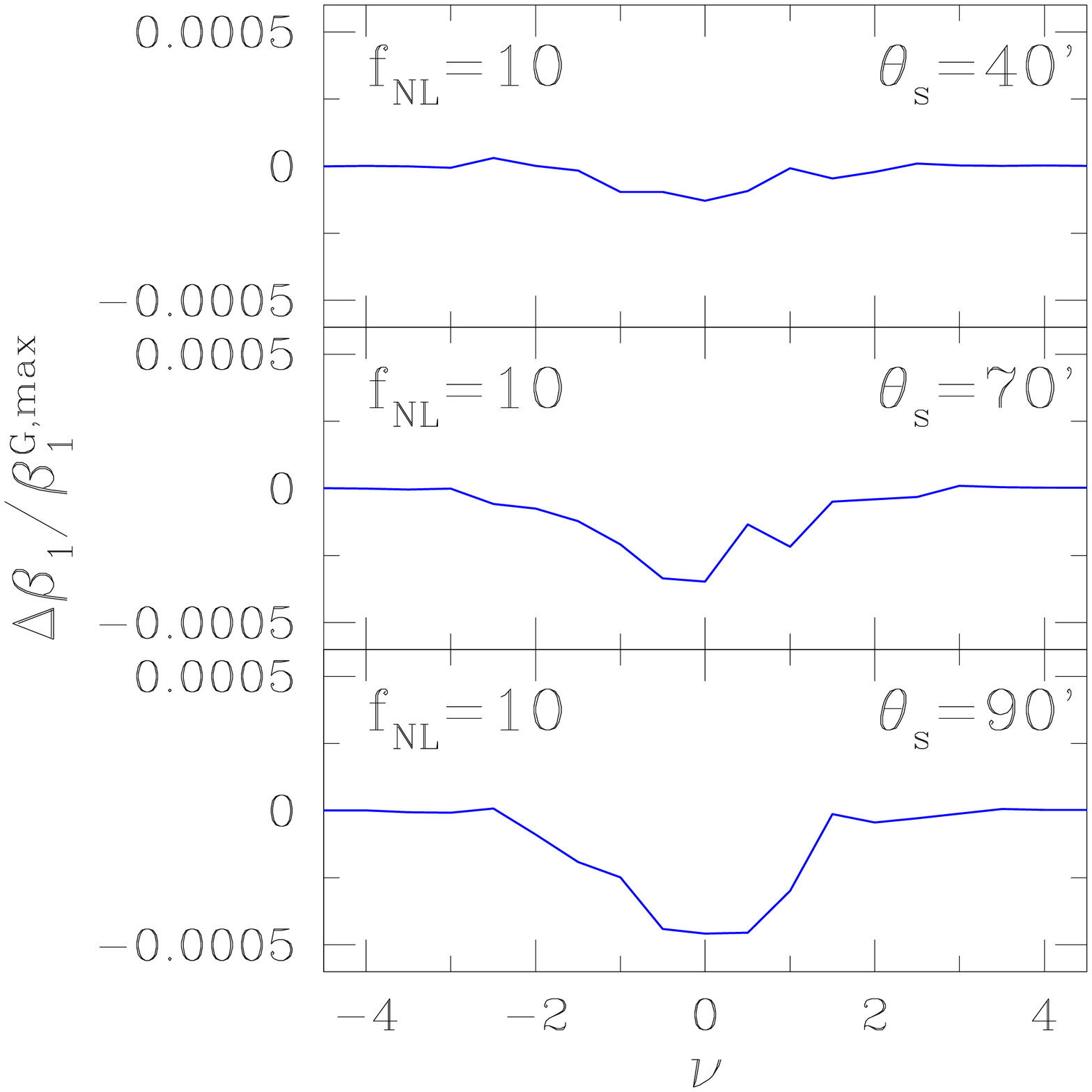}}
\end{center}
\caption{Betti numbers and their non-Gaussian deviations for $\tilde I$, for $\fnl=10$ and different smoothing angles. }
\label{fig:bettiintensity}
\end{figure}

\begin{figure}[h]
\begin{center}
\resizebox{1.8in}{1.8in}{\includegraphics{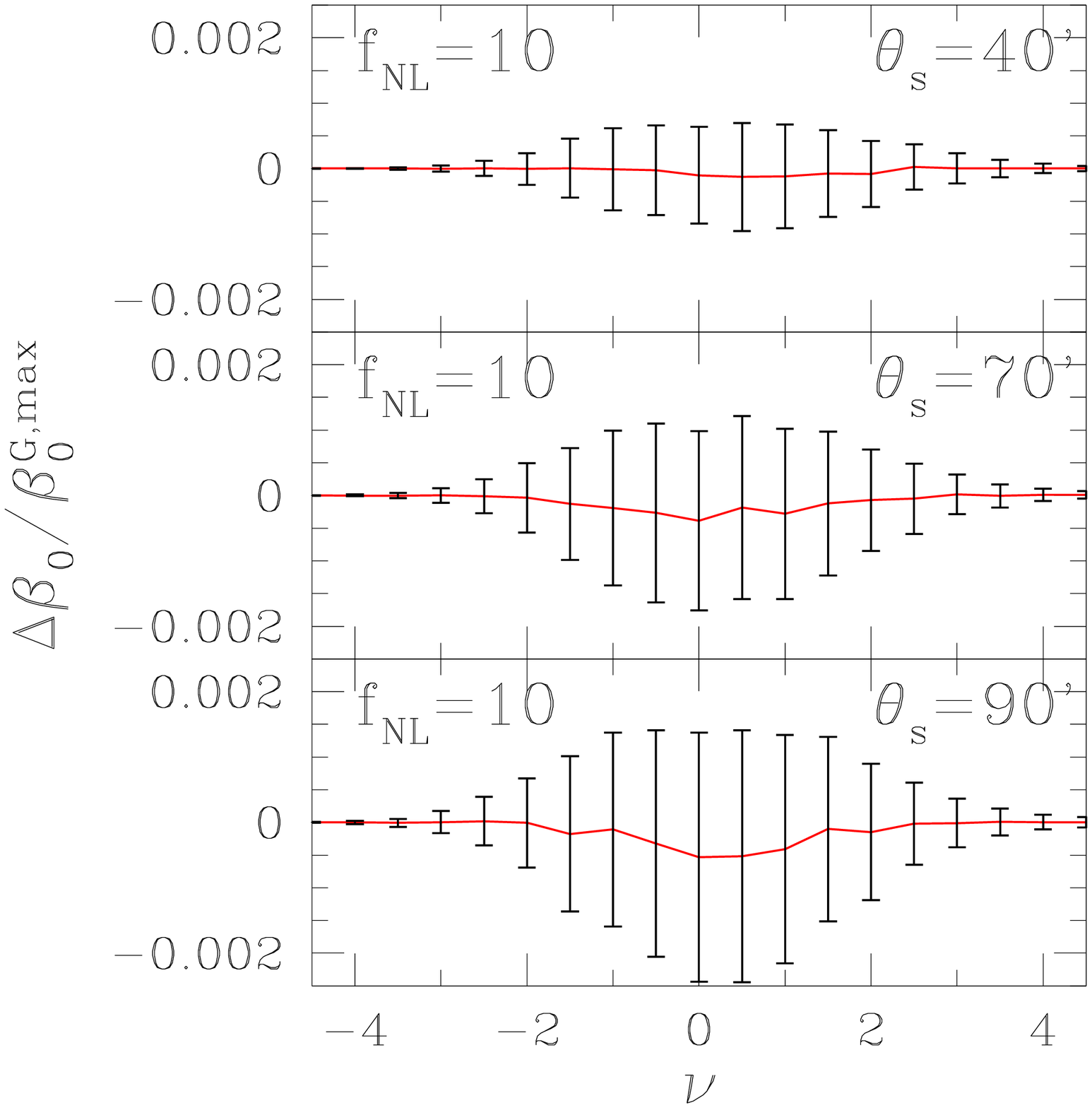}}
\resizebox{1.8in}{1.8in}{\includegraphics{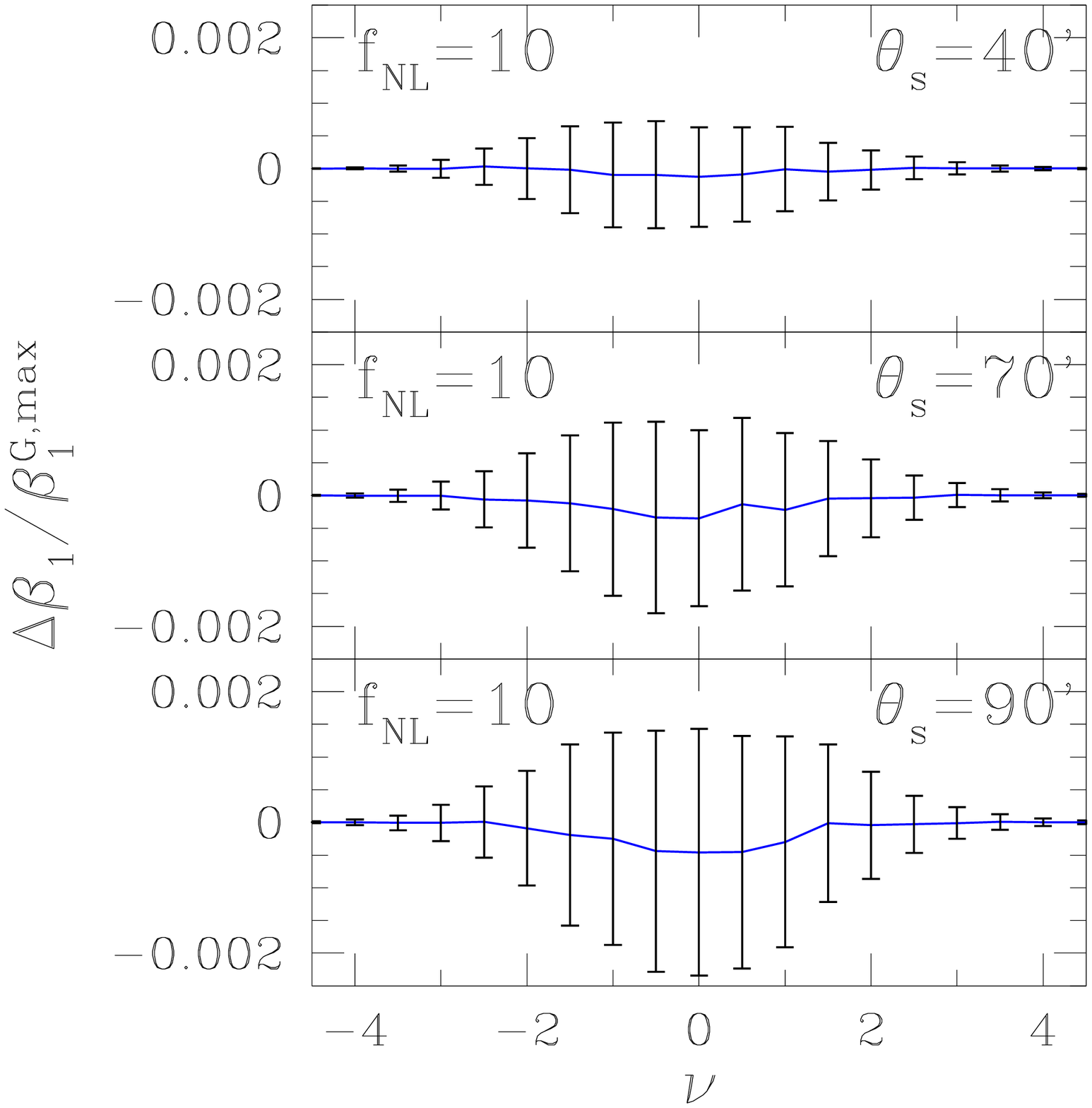}}
\end{center}
\caption{Non-Gaussian deviations of Betti numbers for $\tilde I$ with
  error bars for the same $\fnl$ and smoothing angles as
  Fig.~(\ref{fig:bettiintensity}). }
\label{fig:bettiintensity_errorbar}
\end{figure}

Fig.~(\ref{fig:bettiintensity}) shows Gaussian $\beta_0$ and $\beta_1$
(left panel) and their respective non-Gaussian deviations (middle and
right panels) for $\tilde I$, averaged over 1000 simulations. The
amplitudes of Gaussian $\beta_0$ and $\beta_1$ are comparable to those
of $E$. The trend for the amplitudes of non-Gaussian deviations is
also similar. They are much smaller for small values of $\theta_s$,
and increases as $\theta_s$ increases. Fig.~(\ref{fig:bettiintensity_errorbar}) shows the same deviations as Fig.~(\ref{fig:bettiintensity}) but on a larger scale in order to highlight the large error bars. This again implies lower
statistical power for $\tilde I$ to detect local type non-Gaussianity.


\subsection{Comparison of statistical sensitivity of $T$, $E$ and $\tilde I$ to primordial non-Gaussianity}

A simple way of estimating the statistical non-Gaussian information
encoded in temperature fluctuations, $E$ and $\tilde I$ and comparing
them, is to integrate the absolute values of the non-Gaussian
deviations measured in units of the corresponding sample variances
from $\nu=-4.5$ to 4.5. Let us define the quantity $A$ as,
\begin{equation}
A=\Delta\nu\sum_{i=1}^M\left(|\Delta O(i)|/O^{\rm G,max}\right)/\sigma_s(i),
\label{eqn:area}
\end{equation} 
where $M$ is the number of threshold levels with spacing $\Delta\nu$,
$O$ can be either $V_1$, $V_2$, $\beta_0$ or $\beta_1$, and
$\sigma_s(i)$ are the respective sample variances at each threshold
level $i$. For Mfs we have $M=13$. 

The resulting values are shown in Table (\ref{table:area}). We find
comparable values of $A$ for all four observables (excluding the area
fraction) for temperature fluctuations and $E$. Therefore, we conclude
that $E$ can be as useful as temperature fluctuations data for
constraining primordial non-Gaussianity. The values of $\tilde I$ are
however much lower and hence it lacks statistical power for such
analysis.

\begin{center}
\begin{table}
\begin{center}
\begin{tabular}{|c|c|c|c|c|c|}
\hline
Field & Smoothing angle & $A$ for $V_1$ & $A$ for $V_2$ \\
\hline
 &$40'$ &  27.3   & 22.8   \\
 $\Delta T/T$ & $70'$ &  21.3   & 14.5   \\
& $90'$ &  18.4   & 26.0   \\
\hline
\hline
 &$40'$ &  34.8   & 26.0   \\
 $E$ & $70'$ &  19.4   & 19.0   \\
& $90'$ &  14.0   & 17.0   \\
\hline
\hline
 & $40'$ &  1.2   & 1.2   \\
 $\tilde I$ & $70'$ &  1.8   & 1.6 \\
& $90'$ &  1.7   & 1.5  \\
\hline

\end{tabular}
\end{center}
\caption{Table showing values of $A$ defined in Eq.~(\ref{eqn:area})
  for $V_1$, $V_2$, $\beta_0$ and $\beta_1$. }
\label{table:area}
\end{table}
\end{center}

\section{Conclusion}

In this paper we have investigated the signatures of local type
primordial non-Gaussian scalar perturbations on CMB polarization. Such
a study has become very important with the availability of
increasingly accurate polarization data. To this end we have used
local type non-Gaussian simulations of $E$ mode polarization and the
total polarization intensity $I$ and calculated non-Gaussian
deviations of Minkowski functionals and Betti numbers. Since in our
case $I$ is constructed from simulated $E$ mode maps, the non-Gaussian
information contained in it is essentially the same as in $E$.  In
realistic observational situation $I$ is expected to have
contributions from $B$ modes and different systematic instrumental
errors can have different signatures in $E$, $B$ and $I$. Hence we
must analyze all three fields or possibly other clever constructs from
them to isolate different sources of non-Gaussianity or check for
consistency in the data.

We first derive the expected shapes of PDFs of non-Gaussian $E$ mode and
$I$ fields. $E$ mode has deviations of order $\fnl\sigma$, similar to that
of temperature fluctuations, as expected. $I$, on the other hand, has
deviations proportional to $(\fnl\sigma)^2$ at the lowest order. Calculations of the PDF and its non-Gaussian deviation from simulated maps for $I$ and confirmed smallness of the amplitude of the non-Gaussian deviation in comparison to that of $E$. We found that the
statistical fluctuations are larger. This implies that $I$
will not be a good field to use for the detection of local-type non-Gaussianity. 

Next we analyze the geometrical and topological properties of
excursion sets of $E$ and $I$ using Minkowski Functionals and Betti
numbers. We find that the non-Gaussian deviations of all the
observables for $E$ field is similar in shape, amplitude and size of
error bars to that of temperature fluctuations. We conclude that $E$
modes will provide independent and equally strong constraint on
$\fnl$. Non-Gaussian deviations of $I$ are much smaller and the error
bars are much larger. This is in agreement with what we had obtained
for the PDF. So this further implies that $I$ by itself will not
constrain $\fnl$ well. However, when used in conjunction with $E$
modes, information from $I$ can be extremely valuable in
distinguishing different types of non-Gaussianity. For example if we
measure the non-Gaussian deviations of MFs from observational data and
find that $E$ and $I$ have similar levels of deviations then it will
strongly indicate that the source of non-Gaussianity is not $\fnl$
type.

In this work we use the clean CMB signal maps and ignored the effects of instrumental noise, beam
shape and incomplete sky. It must be noted that instrumental noise
levels for polarization in observations such as WMAP are larger than
that of temperature fluctuations. This will downgrade the power of polarization map in comparison to the temperature map in constraining $\fnl$ in real situations. We are currently investigating this issue. Inclusion of
$B$ mode will modify the properties of the total polarization
intensity. Recently data from BICEP2~\cite{Ade:2014xna} has indicated
a rather large value for the tensor-scalar ratio, which translates
into a relatively large rms value for $B$ field. Further $B$ modes can
carry non-Gaussian information of the primordial tensor
perturbations. We are studying these cases in ongoing work and plan to
apply the analysis to observational data from PLANCK and other future experiments.

\ack{
The computation required for this work was carried out on the Hydra
cluster at IIA. We acknowledge use of the HEALPIX
package~\cite{Gorski:2005, Healpix} which was used to derive some of
the results. We thank F.~Elsner and B.~Wandelt for the use of their
simulations. P.~C.~ would like to thank Sreedhar B. Dutta for useful
discussions.
\appendix
\section{PDF of the total polarization intensity in the presence of $B$ mode} 

To see the effect of inclusion of $B$ mode on the PDF of $I$ we have plotted it in Fig.~(\ref{fig:pdf_I_r}) for input Gaussian $E$ and $B$ with tensor-to-scalar ratio $r=0.2$, for the same smoothing angles as in Fig.~(\ref{fig:pdf_I}). The smoothing was done on $E$ and $B$ maps. When $B$ mode is included we find that the peak amplitude of the PDF is lower and this becomes visible at larger smoothing angles. 
\begin{figure}[h]
\begin{center}
\resizebox{2.5in}{2.in}{\includegraphics{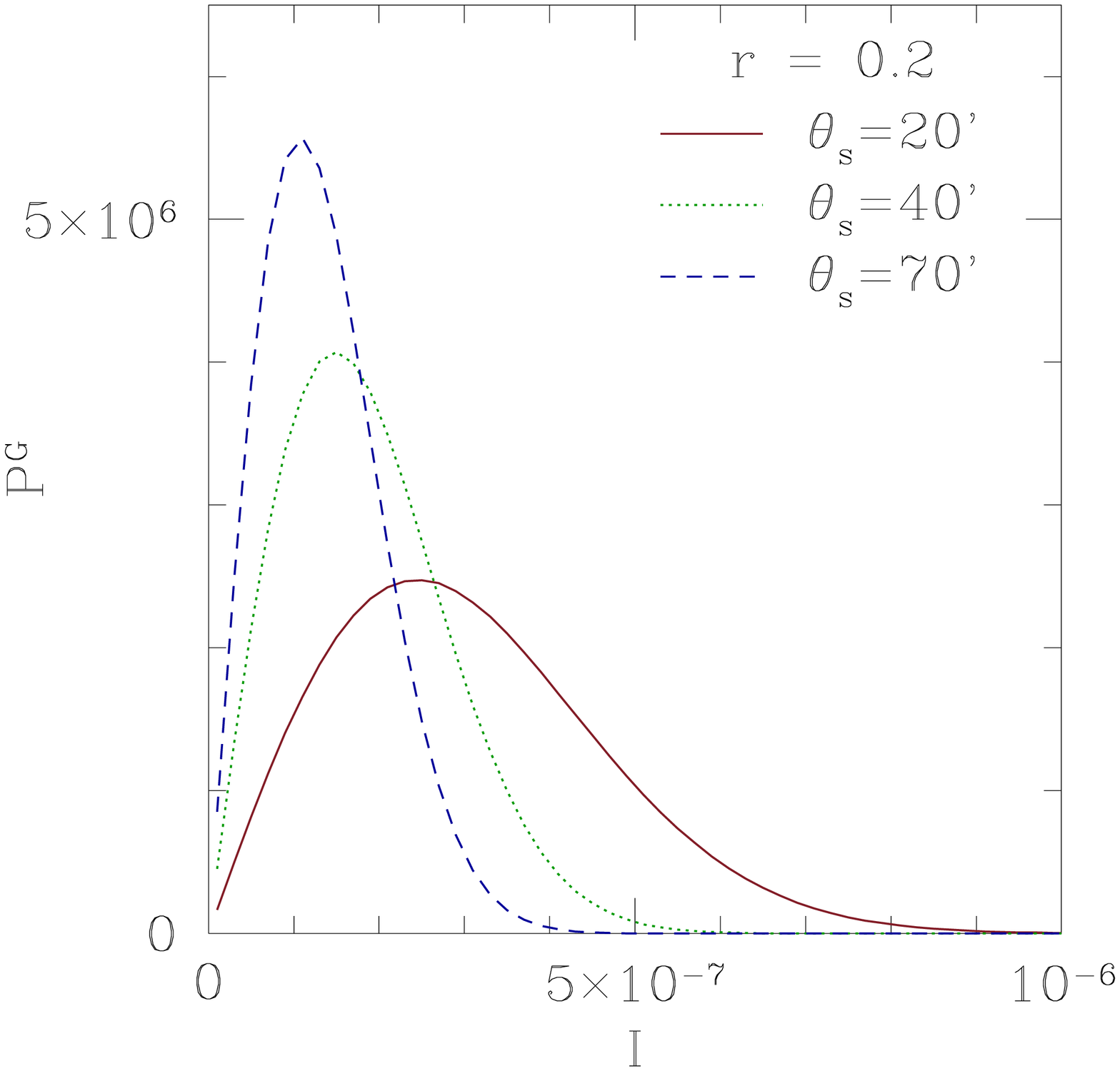}}
\end{center}
\caption{ PDF of $I$ for input Gaussian $E$ and $B$ with tensor-to-scalar ratio value given as $r=0.2$. Smoothing was done on  $E$ and $B$ maps. Plots are average over 1000 simulations.}
\label{fig:pdf_I_r}
\end{figure}


\end{document}